\documentstyle[eqsecnum,preprint,aps]{revtex}
\begin{document}
\draft
\title{COSMOLOGICAL INFLATION, \\
       MICROWAVE BACKGROUND ANISOTROPIES AND \\
       LARGE SCALE STRUCTURE OF THE UNIVERSE}

\author{D. S. Salopek}
\vspace{2pc}
\address{
Department of Physics, University of Alberta,
Edmonton, Canada T6G 2J1 }
\vspace{2pc}

\maketitle

\begin{abstract}
Cosmological inflation provides the simplest and most
promising mechanism for generating fluctuations for
structure formation. Using powerful Hamilton-Jacobi methods,
I will describe (1) how to  compute density fluctuations
and cosmic microwave anisotropies arising from inflation, and
(2) improvements of the Zel'dovich
approximation describing gravitational collapse. I compare
these results with the latest cosmological observations.

\noindent
{\bf Keywords:} cosmology / inflation / large scale structure of the
Universe/ microwave background anisotropies / semiclassical gravity

\end{abstract}
\vspace{2pc}

\vspace{2pc}
\begin{center}
{\it in Proceedings of the }\\
International School of Astrophysics ``D. Chalonge'' \\
Third Course: Current Topics in Astrofundamental Physics \\
Erice, Italy, September 4-16, 1994 \\
Edited by N. Sanchez (Kluwer Academic Publishers)
\end{center}

\section{INTRODUCTION AND REVIEW}

The field of cosmology remains one of the
most exciting areas in physics. A new branch of
astronomy was recently opened with the detection of the
cosmic microwave background anisotropy in 1992 by the
Cosmic Background Explorer (COBE) satellite \cite{WRIGHT92}.
Of equal importance to astronomy is  the tremendous
growth in (1) galaxy redshift surveys and (2) measurements of
intermediate-angle microwave anisotropies. Although one cannot
predict the precise outcome, the impact in theoretical cosmology
should be quite pronounced.


The inflationary scenario is perhaps the most promising
mechanism for explaining microwave background fluctuations
as well as fluctuations for galaxy formation.
Before discussing theoretical strategies, I will first
review the observational clues that suggest that
we are on the right track.

  \subsection{Observational Clues Supporting Inflation }

    \subsubsection{Critical Density of the Universe}

A galaxy survey compiled by the Infrared Astronomical
Satellite (IRAS) can be employed in conjunction with various
redshift surveys to compute the mean density of the Universe.
One finds \cite{PD94}
\begin{equation}
\Omega^{0.6}/b_{IRAS}= 1.0 \pm 0.2 \, \label{CRIT}
\end{equation}
where $\Omega \equiv \rho / \rho_{crit}$ and
$b_{IRAS}$ is the biasing parameter for
IRAS galaxies. Provided that  $b_{IRAS} \sim 1$
(which is quite reasonable to assume), it is consistent
to assume that the Universe is at critical density
which is the strongest prediction of inflation.

    \subsubsection{Large Angle Microwave Anisotropies}

Using statistical arguments in 1992, the COBE team announced
that they had detected a microwave anisotropy using one year
of DMR (Differential Microwave Radiometer) data \cite{WRIGHT92}.
Soon after, an MIT group confirmed their results \cite{GANGA93}.
The Tenerife experiment was actually the first to measure the
anisotropy in one patch of the sky \cite{LAS94}.

With two years of data, the COBE team determined that the
spectral index for scalar perturbations, which describes the
slope of the power spectrum,  was $n= 1.17 \pm 0.31$
\cite{BENNETT94}.
The simplest inflation models give $n < 1$. In this limited
sense inflation is consistent with COBE.

    \subsubsection{Hubble Parameter?}

Is the present value of the Hubble parameter $H_0$ consistent with
inflation? Unfortunately, this is still an open question.
In order for the age, $t_0=2/(3 H_0)$, of an Einstein-deSitter
universe to be larger than that of the oldest stars, one
requires that $H_0 \leq 50 \; {\rm km \; s}^{-1} {\rm Mpc}^{-1}$.
At this conference, the Hubble parameter controversy was
reviewed by Tammann \cite{TAMMANN94}. Personally, I prefer a new technique
utilizing the Sunyaev-Zel'dovich (SZ) effect which describes
how hot gas within a cluster of galaxies produces
a temperature deficit $\Delta T_{SZ}$ in the
cosmic microwave background. The SZ effect has been
observed by Birkinshaw {\it et al} \cite{B91} and
by  a group at Cavendish in Cambridge \cite{LAS94}.
Throughout this article, I will assume that
$H_0 = 50 \; {\rm km \; s}^{-1} {\rm Mpc}^{-1}$,
which is consistent with measurements of the SZ effect.

    \subsubsection{Structure Formation?}

Is the observed pattern of galaxies consistent with inflation?
Within an order of magnitude, the answer is yes.
However, much attention has been given to the fact that
the observations deviate from the standard
cold-dark-matter (CDM) model \cite{PEEBLES82}
with an $n=1$ spectrum by
a factor of about three over about a decad in wavenumber.
My own view is that the discrepancy is not very serious
and that higher order nonlinear corrections, slight variations in
the models, increases in the amount of data, {\it etc.}, will
resolve this issue. I believe that most of the
basic ingredients have been correctly identified
and that wholesale modifications such
as abandoning inflation, invoking defect models, {\it etc.},
are probably not necessary.

For example, there are many inflationary models, and
one now wishes to resolve which is the correct one.
Here I will focus on one class of models which is particularly
easy to deal with: power-law inflation. These models
yield substantial amounts of primordial gravitational
waves and they also give rise to more power at large scales.
They alleviate the problems of large scale structure.

As the data for galaxy clustering and microwave background
anisotropies (both intermediate  and large angle)
continue to improve over the next few years,
we will be better able to identify the correct inflationary
model. In this way, it is possible to probe the earliest
epochs of the Universe \cite{S92}.

\subsection{Theoretical Strategy}

Faced a with an increasing number of observations,
it proves extremely useful to have a theoretical formalism that
encompasses virtually all aspects of cosmology. In this
way, theory attempts to put some order in what otherwise
looks like chaos. I strongly advocate using the Hamilton-Jacobi formalism
for general relativity \cite{SS94}-\cite{SB1}.
The reasons are quite are numerous:

\noindent
{\bf (1) Inflation requires a Quantum Theory of Gravity.}
However, the full theory does not exist. Since it is generally believed
that tensor fluctuations arising from inflation began initially in the
ground state \cite{S92},
it is absolutely essential to quantize the gravitational
field. At the present, one should perhaps be content with a semi-classical
approximation where one approximates the wavefunctional by
\begin{equation}
\Psi \sim e^{i {\cal S} / \hbar } \, , \label{sc}
\end{equation}
where $\hbar$ is Planck's constant.
The phase factor ${\cal S}$ is often referred to as the generating
functional. It satisfies the Hamilton-Jacobi
equation for general relativity. If ${\cal S}$ where real, then
this approximation would be entirely classical. However, by
allowing ${\cal S}$ to be complex, one can indeed describe
quantum phenomena, including fluctuations beginning in the
ground state.

\noindent
{\bf (2) Solving the Constraint Equations of General Relativity.}
For example, the momentum constraint associated with the $G^0_i$
Einstein equation is rather trivial to satisfy, since
it legislates that the generating functional be invariant
under reparametrizations of the spatial coordinates.

\noindent
{\bf (3) Covariant Description of the Gravitational Field.}
Using HJ theory, one can perform hypersurface-invariant
computations. In this sense, one can resolve the problem of time:
time is arbitrary. Moreover, the HJ formalism
includes almost everything that you wanted to know about
gauge-invariant variables but were afraid to ask.
For an alternative approach describing covariant perturbations
of general relativity, see ref.\cite{ELLIS94}.

\noindent
{\bf (4) Finding Exact Solutions.}
HJ theory is useful in deriving exact solutions
for cosmological situations of physical interest:
during inflation, during the radiation/matter dominated
epochs, etc.. Here I will review exact solutions of the
long-wavelength equations and the perturbation equations for
power-law inflation.

\noindent
{\bf (5) Applying the Zel'dovich Approximation to
General Relativity.}

HJ methods allow one to discuss the Zel'dovich approximation
\cite{Zel} in a relativistic framework \cite{CPSS94},
\cite{SSC94}, \cite{SSCP94}. The Zel'dovich approximation
describes the formation of sheet-like structures which
appear to be quite common in our Universe.

In Sec. 2, I will describe the most useful data sets available
to cosmologists. Hamilton-Jacobi theory will be
reviewed in Sec. 3. Several hypersurface-invariant
approximation schemes will be discussed including
the {\it spatial gradient expansion} and the
{\it quadratic curvature approximation}. In Sec. 4, I
discuss the phenomenological consequences
of various inflation models. Applications of HJ theory to the
Zel'dovich approximation are given in Sec. 5.
Attempts to go beyond the semi-classical approximation
using one-loop calculations will be summarized in Sec. 6.
Finally, I conclude in Sec. 7.

(Unless otherwise stated, units will be chosen so that
$c=k= 8 \pi G = 8 \pi/ m^2_{\cal P}= \hbar=1$.)

\section{DESCRIPTION OF COSMOLOGICAL DATA}

For a quantitative analysis, the two best cosmological data sets are:
(1) the large angle ($> 3^0$) microwave anisotropies
determined by COBE and (2) galaxy correlation functions.
A third data set,
(3) microwave background measurements at intermediate angles,
shows wide variations, and it presently does not set useful quantitative
limits.

    \subsection{Large Angle Microwave Anisotropy}

In Fig.(1), I show the
correlation function in the temperature
\begin{equation}
C(\alpha) = < \Delta T(x) \Delta T(x^\prime) >
\end{equation}
with the monopole and dipole removed. It is
computed using the two-year COBE data set.
For angles less than $60^0$, the error bars are much smaller than
than the values of measurements, and one is impressed by the precision
of the experiment. However when one factors out all the
enthusiasm and publicity, there are basically two numbers that follow
from this graph: the amplitude and the slope.

The slope is often given in terms of the spectral index,
which they determine to be  $n=1.17 \pm 0.31$ \cite{BENNETT94}.
In their initial analysis, they included the quadrupole component after
subtracting the contribution of the Milky Way.
Since this subtraction is quite tricky and since
cosmic variance for the quadrupole is large, one may wish to
repeat the calculations without including the quadrupole, in which
case one finds $n = 0.96 \pm 0.36$. In either case, their
results provide a test of inflation, and they are consistent with
the simplest models of inflation which yield $n < 1$.

The amplitude is given in terms of the temperature anisotropy
\begin{equation}
\sigma_{sky}(10^0) = 30.5 \pm 2.7 \mu K  \label{NORM}
\end{equation}
for a Gaussian beam of $10^0$ FWHM (full-width-at-half-maximum).
This value does not differ much from the first-year results,
but the $10\%$ error bars represent an improvement by
a factor of two. When one fits various inflations,
I will disregard COBE's determination of the spectral
index, and use only $\sigma_{sky}(10^0)$ since
it has the smallest error limits.

    \subsection{Large Scale Structure of the Universe}

Over the past five years, there has been tremendous
improvement in the quantity and quality of
galaxy clustering data. Frenk \cite{F94} has given an
excellent review of the situation at this conference.
Here I will touch on the main points.

Recently, Peacock and Dodds \cite{PD94} have
compiled the most useful galaxy data in a
user-friendly form. They consider eight
data sets, including: one angular survey,
(1) the APM (Automatic Plate Machine)
and seven redshift surveys (2) Abell clusters, (3) Radio
galaxies, (4) Abell vs. IRAS (Infrared galaxies),
(5) CfA (Center for Astrophysics),
(6) APM/Stromlo, (7) Radio vs. IRAS
and (8) IRAS. Corrections are made for
nonlinear evolution and for redshift space distortions.
In addition, before they merge the data, they rescale
each set using a biasing parameter $b_{(i)}, i=1,...,8$.
A redshift space correction implies that the
IRAS biasing parameter is constrained by eq.(\ref{CRIT}).
If we accept the inflationary model with
$\Omega =1$, we see that it is consistent to
assume that the biasing parameter $b_{IRAS}$ for IRAS
galaxies is unity. Hence, from now on we will assume that
that IRAS galaxies trace the mass:
$b_{IRAS}= 1$. I wish to emphasize that the
biasing parameter is no longer a free parameter.

Using all data sets, they
determine the {\it linear} power spectrum
\begin{equation}
{ \cal P}_\rho(k) \equiv { k^3 \over 2 \pi^2 } \,
\int d^3x \, e^{-ik \cdot x} \; < { \delta \rho(x)  \over \rho}
{ \delta\rho (0)  \over \rho} >
\end{equation}
for the density perturbation $\delta \rho$ at the present epoch.
It is $k^3/(2 \pi^2)$ multiplied by the Fourier transform
of the two-point correlation function in the density.
It is plotted in Fig.(2). The precision is quite
remarkable, considering how much scatter there
was in each of the individual data sets.
For comparison, the 1.8-power-law
usually associated with optical galaxies is also shown.
At smaller scales where the 1.8-power-law is valid,
optical galaxies are definitely biased tracers.

    \subsubsection{Intermediate Angle Anisotropies}

Wright {\it et al} \cite{INTERMED4} have compiled a list
of microwave background anisotropy measurements. Their results are
plotted in Fig.(3). The solid line depicts the
standard prediction of $n=1$ CDM with no gravitational
wave contribution. Unfortunately, there is much
scatter in the measurements for $\ell>20$. In addition, one suspects
that there is contamination from foreground objects.

Although this is a promising line of research
(for a review, see ref.\cite{SMOOT94}),
the present quality of intermediate-angle data does
not lead to any firm quantitative conclusions.
For this reason, when considering fits of inflationary models,
I will utilize only  the (1) large
angle microwave normalization and (2) galaxy clustering data.

\section{HAMILTON-JACOBI THEORY}

Theoretical computations of various inflation models
can be described in very elegant terms using
Hamilton-Jacobi methods. Some very deep theoretical
issues can be addressed.
Previously, there had been some confusion in the
choice of gauge. For example, what should one
choose for the time hypersurface? Other problems include
choosing the initial wavefunctional.

\subsection{Resolving the Question of Time for Semi-Classical Gravity}

Before getting into intricacies of HJ theory, I would
like to demonstrate how it is possible to illuminate
the role of time in semi-classical gravity. A simple
analogy from potential theory illustrates the basic
point which is easy to understand.

\subsubsection{Potential Theory}

The fundamental problem in potential theory is: given a force
field $g^i(u_k)$ which is a function of $n$ variables $u_k$,
what is the potential $\Phi \equiv \Phi(u_k)$ (if it exists)
whose gradient returns the force field,
\begin{equation}
{\partial \Phi \over \partial u_i} = g^i(u_k) \quad ?
\end{equation}
Not all force fields are derivable from
a potential. Provided that the force field satisfies the
integrability relation,
\begin{equation}
0= {\partial g^i \over \partial u_j} - {\partial g^j \over \partial u_i} =
\left [{\partial  \over \partial u_j}, {\partial  \over  \partial u_i }
\right ] \, \Phi \, , \label{CONSIST}
\end{equation}
(i.e., it is curl-free),
one may find a solution which is conveniently expressed using a
line-integral
\begin{equation}
\Phi(u_k) = \int_C \sum_j dv_j \ g^j(v_l) \ .
\end{equation}
If the two endpoints are fixed, all contours return the same
answer. In practice, I will employ the simplest contour that
one can imagine: a line connecting the origin to the
observation point $u_k$. Using $s$, $0 \le s \le 1$,
to parameterize the contour, the line-integral may be rewritten as
\begin{equation}
\Phi(u_k) = \sum_{j=1}^n \int_0^1  ds  \; u_j \ g^j(su_k) \ .
\label{lint}
\end{equation}

Infinite dimensional line integrals appear in solutions of
the Hamilton-Jacobi equation for general relativity.
Basically, each contour corresponds to a specific choice
of time-hypersurface. They all yield the same answer
provided certain consistency relations analogous to
eq.(\ref{CONSIST}) are met.

\subsection{Hamilton-Jacobi Equation for General Relativity}

The Hamilton-Jacobi equation for general relativity is
derived using a Hamiltonian formulation of gravity.
One first writes the line element using the ADM 3+1 split,
\begin{equation}
ds^2=\left(-N^2+\gamma^{ij}N_i N_j\right)dt^2 +
2N_idt \,dx^i + \gamma_{ij}dx^i\,\ dx^j\ ,
\label{ADMdecomp}
\end{equation}
where $N$ and $N_i$ are the lapse and shift functions, respectively,
and $\gamma_{ij}$ is the 3-metric. Hilbert's action for gravity interacting
with a scalar field becomes
\begin{equation}
{\cal I}=\int d^4x\left(\pi^{\phi}\dot\phi +\pi^{ij}\dot\gamma_{ij}
-N{\cal H} -N^i{\cal H}_i\right).
\label{ADMaction}
\end{equation}
The lapse and shift functions are Lagrange multipliers that
ensure that the energy constraint ${\cal H}(x)$ and the
momentum constraint ${\cal H}_i(x)$ vanish.

The object of chief importance is the generating functional
\begin{equation}
{\cal S}\equiv {\cal S}[\gamma_{ij}(x), \phi(x)].
\end{equation}
For each universe with field configuration
$[\gamma_{ij}(x), \phi(x)]$ it assigns a number
which can be complex. The generating functional is
the `phase' of the wavefunctional in the semi-classical approximation,
eq.(\ref{sc}).
For the applications that we are considering, the prefactor
before the exponential is not important, although
it has interesting consequences for quantum
cosmology\cite{BK94}. The probability ${\cal P}$
of finding a field configuration
is given by the square of the wavefunctional:
\begin{equation}
{\cal P} \equiv |\Psi|^2 \, .
\end{equation}

Replacing the conjugate momenta by functional derivatives
of ${\cal S}$ with respect to the fields,
\begin{equation}
\pi^{ij}(x)={\delta{\cal S}\over \delta{\gamma_{ij}(x)}}\ , \qquad
\pi^{\phi}(x)={\delta{\cal S}\over \delta\phi (x)}\ ,
\label{pis}
\end{equation}
and substituting into the energy constraint equation,
one obtains the Hamilton-Jacobi equation,
\begin{eqnarray}
{\cal H}(x)=&&\gamma^{-1/2} {\delta{\cal S}\over \delta\gamma_{ij}(x)}
{\delta{\cal S}\over \delta\gamma_{kl}(x)}
\left[2\gamma_{il}(x) \gamma_{jk}(x) - \gamma_{ij}(x)\gamma_{kl}(x)\right]
\nonumber \\
&& + {1\over 2} \gamma^{-1/2}
\left({\delta{\cal S}\over \delta\phi(x)}\right)^2
+\gamma^{1/2}V(\phi(x)) \nonumber \\
&& -{1\over 2}\gamma^{1/2}R
+{1\over 2} \gamma^{1/2}\gamma^{ij}\phi_{,i}\phi_{,j}=0 \ ,
\label{HJequation}
\end{eqnarray}
which describes how ${\cal S}$ evolves in superspace.
$R$ is the Ricci scalar associated with the 3-metric, and $V(\phi)$
is the scalar field potential.
In addition, one must also satisfy the momentum constraint
\begin{equation}
{\cal H}_{i}(x)=-2\left(\gamma_{ik}{\delta{\cal S}\over \delta\gamma_{kj}(x)}
\right)_{,j} +
{\delta{\cal S}\over\delta\gamma_{lk}(x)}\gamma_{lk,i} +
{\delta{\cal S}\over\delta\phi (x)} \phi_{,i}=0 \ ,
\label{Smomentum}
\end{equation}
which legislates that ${\cal S}$ be invariant under
reparametrizations of the spatial coordinates.
Since neither the lapse function
nor the shift function appears in
eqs.(\ref{HJequation},\ref{Smomentum}) the temporal
and spatial coordinates are {\it arbitrary}:
HJ theory is {\it covariant}.

\subsection{Spatial Gradient Expansion}

It appears initially that finding solutions to the HJ equation
will be a hopeless task since one is essentially describing an
ensemble of evolving universes. However,
it is not very difficult to obtain approximate
solutions. In the first attempt to solve eq.(\ref{HJequation}),
I will expand the generating functional
\begin{equation}
{\cal S}= {\cal S}^{(0)} + {\cal S}^{(2)} + {\cal S}^{(4)} + \dots\ ,
\label{theexpansion}
\end{equation}
in a series of terms according to
the number of spatial gradients that they contain.
As a result, the Hamilton-Jacobi equation can likewise be
grouped into terms with an even number of spatial
derivatives:
\begin{equation}
{\cal H}={\cal H}^{(0)} + {\cal H}^{(2)} + {\cal H}^{(4)} + \dots\ .
\label{theexpansion2}
\end{equation}
The invariance of the generating functional
under spatial coordinate transformations
(see eq.(\ref{Smomentum})) suggests a solution of the form,
\begin{equation}
{\cal S}^{(0)}[ \gamma_{ij}(x), \phi(x)] = - 2 \int d^3x
\gamma^{1/2} H \left[ \phi(x) \right] \ , \label{LW}
\end{equation}
for the zeroth order term ${\cal S}^{ (0) }$. The function
$H \equiv H(\phi)$ satisfies the HJ equation of order zero,
\begin{equation}
H^2={2\over 3}\left( {\partial H\over\partial\phi} \right)^2
 +{ V\left( \phi \right ) \over 3}  \ ,
\label{Hequation}
\end{equation}
which is an nonlinear ordinary differential equation.
Note that ${\cal S}^{(0)}$ contains no spatial gradients.
It describes the evolution of long-wavelength fields----
these are inhomogeneous field configurations where
the spatial gradients of the inhomogeneities are so small
that they do not alter the evolution significantly from
that of a homogeneous model. In fact, long-wavelength fields
are an essential feature of all structure formation
scenarios arising in cosmology.

We will examine in detail the special
case of inflation with an exponential potential \cite{LM85}
\begin{equation}
V(\phi)=V_0 \, \exp\left(-\sqrt{2\over p}\phi\right),
\label{s.f.potential}
\end{equation}
which has the exact solution
\begin{equation}
H(\phi)= \left[{V_0\over 3 \left ( 1-1/(3p) \right)}\right]^{1/2}
\exp\left(-{\phi\over\sqrt{2p}}\right) \, .
\qquad {\rm (single\ scalar\ field)}
\label{zero.scalar}
\end{equation}
It yields power-law inflation where the scale factor
evolves according to $a(t) \propto t^p$.

In order to compute higher order terms in the spatial
gradient expansion, one introduces a
change of variables, $( \gamma_{ij}, \phi ) \rightarrow
(f_{ij}, u)$:
\begin{equation}
u = \int \  { d \phi \over -2 { \partial H \over \partial \phi } } \ ,
\quad f_{ij} = \Omega^{-2}(u) \, \gamma_{ij} \ , \label{changeA}
\end{equation}
where the conformal factor $\Omega \equiv \Omega(u)$ is defined through
\begin{equation}
{ d \ln \Omega \over d u} \equiv -2 { \partial H \over \partial \phi}
{ \partial \ln \Omega \over \partial \phi} = H \ . \label{changeB}
\end{equation}
in which case the equation for ${\cal S}^{(2m)}$ becomes
\begin{equation}
{\delta {\cal S}^{(2m)}\over\delta u(x)}\Bigg|_{f_{ij}}
+ {\cal R}^{(2m) }[u(x), f_{ij}(x)] =0\ .
\label{HJ.conf.for.dust}
\end{equation}
The remainder term ${\cal R}^{(2m) }$ depends on some quadratic
combination of the previous order terms (it may be written explicitly).
For example, for $m=1$, it is
\begin{equation}
{\cal R}^{(2)} =
{1\over 2} \gamma^{1/2}\gamma^{ij}\phi_{,i}\phi_{,j}
-{1\over 2}\gamma^{1/2}R \, .
\end{equation}
Eq.(\ref{HJ.conf.for.dust}) has the form of an infinite
dimensional gradient. It may integrated using a line
integral analogous to eq.(\ref{lint}):
\begin{equation}
{\cal S}^{(2m)}=-\int d^3x\int_0^1  ds\  u(x) \
{\cal R}^{(2m)}[su(x), f_{ij}(x)] \; ;
\label{lints}
\end{equation}
the conformal 3-metric $f_{ij}(x)$ is held constant
during the integration which may be performed explicitly
in many cases of interest. For example, at second order
in spatial gradients, one finds that
\begin{equation}
{\cal S}^{(2)}[f_{ij}(x), u(x)]=
\int d^3x f^{1/2} \left[ j(u) \widetilde R +
k(u) \, f^{ij} u_{,i} u_{,j}\right] \ . \label{S2A}
\end{equation}
Here $\widetilde R$ is the Ricci curvature of the
conformal 3-metric $f_{ij}$, eq.(\ref{changeA});
the $u$-dependent coefficients $j$ and $k$ are
\begin{equation}
j(u)=\int_0^u {\Omega(u^\prime)\over 2}\;d u'\ + F ,\qquad
k(u)=H(u) \, \Omega(u) \ , \label{S2B}
\label{jkscalar}
\end{equation}
where $F$ is an arbitrary constant.
Provided spatial gauge invariance is maintained,
all contours for the line integral give the same result,
eq.(\ref{S2A}). ${\cal S}^{(2)}$ is useful in
deriving the Zel'dovich approximation from
general relativity (see Sec. 5).

\subsection{Quadratic Curvature Approximation}

In the context of inflation, a finite number of terms in
the spatial gradient expansion is insufficient.
As a result, John Stewart and I have developed a method
where we effectively sum an infinite subset of terms.
We make an Ansatz for the generating functional of the form
\begin{equation}
{\cal S} = {\cal S}^{(0)}+ {\cal S}^{(2)}+ {\cal Q}
\end{equation}
where the functional ${\cal Q}$ is quadratic in the
Ricci tensor of the 3 metric:
\begin{equation}
{ \cal Q} = \int d^3x f^{1/2} \left [
\widetilde R \ \
\widehat S(u, \widetilde D^2) \ \widetilde R +
 \widetilde R^{ij} \ \widehat T(u, \widetilde D^2) \ \widetilde R_{ij}
- { 3 \over 8 }   \widetilde R \ \widehat T(u, \widetilde D^2) \ \widetilde R
\right ] \, .
\label{ansatz2}
\end{equation}
Here $\widehat S(u, \widetilde D^2)$ and $\widehat T(u, \widetilde D^2)$
are the scalar and tensor operators which are functions of $u$
and $\widetilde D^2$.
By using a convenient line-integral in superspace, one
may rewrite the HJ equation in its integral form, which
is useful because it enables one to safely integrate certain
terms by parts. In substituting the
Ansatz, one collects all quadratic terms in the
Ricci curvature, one neglects higher order terms, and finally
one obtains the nonlinear Riccati equations:
\begin{eqnarray}
0&&= { \partial \widehat S \over \partial u} + { 1 \over 8 \Omega^3
\left ( { \partial H \over \partial \phi} \right )^2  } \
\left ( { \Omega \over 2 } - Hj + 8 H \widehat S \widetilde D^2 \right )^2
\, ,
\label{RICCATI1S} \\
0&&= { \partial \widehat T \over \partial u} + { 2 \over  \Omega^3 }
\left ( j + \widehat T \widetilde D^2 \right )^2
\label{RICCATI2S}   \, .
\end{eqnarray}
By making a Riccati transformation, $(S, T) \rightarrow (w, y)$,
these may be reduced to linear equations  for $w$ and $y$:
\begin{eqnarray}
0 =&& { \partial^2 w \over \partial u^2} +
\left \{ 3 H(u) + 2{ \partial \over \partial u} \left [ \ln \left (
{ 1 \over H } { \partial H \over \partial \phi }  \right ) \right ]
\right \}
{ \partial w \over \partial u} - \Omega^{-2}(u) \widetilde D^2 w \, ,
\\ && {\rm (scalar \; perturbations)} \\
\label{scalar} \\
0 =&& { \partial ^2 y \over \partial u^2} + 3 H(u)
{\partial y \over \partial u} -  \Omega^{-2}(u) \widetilde D^2 y \, .
\quad {\rm (tensor\; perturbations)}
\label{tensor}
\end{eqnarray}
For power-law inflation, the solution to both of these
equations may be expressed in terms of Hankel functions.
At early times, far within the Hubble radius, the
solution is chosen to be a positive frequency mode
which describes the ground state
(Bunch-Davies vacuum \cite{BUD78}).
More details are given in ref.\cite{SS94}

\section{PHENOMENOLOGICAL CONSEQUENCES}

Before one can compare theoretical models with
galaxy data, one must make an assumption for the form of
the dark matter. Here I will consider the
simplest cold-dark-matter model \cite{PEEBLES82}.
I will assume a baryon fraction, $\Omega_B = 0.03$, consistent
with nucleosynthesis. In addition, it
is convenient to display the results in Fourier space
where $k$ is the comoving wavenumber.

In Fig.(4), I display several power spectra
for scalar fluctuations,
\begin{equation}
{\cal P}_\zeta(k)  = {\cal P}_\zeta(k_0) \;
\left ( {k \over k_0} \right )^{n-1}   \, , \label{pzeta}
\end{equation}
which may give rise to structure formation. This form
arises from power-law inflation model where
the spectral index for scalar perturbations is given by
\begin{equation}
n=1 - 2/(p-1).
\end{equation}
I have normalized the spectra using COBE's 2-year data set,
eq.(\ref{NORM}). If there were no gravitational
wave contribution to COBE's signal (which is indeed true for n=1),
all the lines would join at $k_0=10^{-4}$Mpc$^{-1}$, the
comoving scale effectively probed by COBE.  Actually,
as $n$ decreases, the gravitational wave contribution
increases, leaving less fluctuations available for the
scalar perturbations: ${\cal P}_\zeta(k_0)$ decreases
as $n$ decreases.

The variable $\zeta$ has proven to be very useful since
it is independent of time provided the wavelength of a mode
exceeds the Hubble radius. This is true during the
inflationary epoch, during heating of the Universe
and even during the radiation and matter dominated
\cite{SBB89},
provided the wavelength of the fluctuations
exceeds the Hubble radius, $ k/ (H a)  << 1$.
In the matter-dominated era,
Bardeen's gauge-invariant variable, $\Phi_H$, and the
density perturbation, $\delta \rho/ \rho$ are related to
$\zeta$ through
\begin{equation}
\Phi_H(k) =  T(k) \, \zeta(k) / 5  \, ,
\end{equation}
\begin{equation}
\delta \rho(k)/ \rho  = { k^2 \tau^2 \over 30}  \;
T(k) \, \zeta(k) \, ,
\end{equation}
where $T(k)$ is the cold-dark-matter transfer function
and $\tau$ is conformal time $ \tau= \int dt/ a(t)$.

Power spectra for the linear density perturbation are
shown in Fig.(5). One sees that cold-dark-matter with $n=1$
is not consistent with the observed data for
$k > 10^{-1.6}$ Mpc$^{-1}$ (short distances), and that the discrepancy
is of the order of a factor of three.

It is useful to
note that the model with $n=0.8$ yields a $50\%$
gravitational wave contribution to $\sigma^2_{sky}(10^0)$.
Looking at Fig.(5), we see that it too is ruled out.
For this reason, we can state: for power-law inflation,
at most $50\%$ of COBE's signal can be the result of
gravitational waves.

For $n=0.9$, gravitational waves contribute $35\%$ to COBE's
signal. This model yields an improved fit, although it is
not quite perfect either since it underestimates the data
by a small factor (less than two) for $k < 10^{-1.6}$ Mpc$^{-1}$
(large distances). This discrepancy is the central problem in
large scale structure.
However, I do not feel that it is very severe for $n=0.9$,
and I will wait for more data before suggesting radical
alterations to the cosmology. In fact, one can further
alleviate the problem by adopting a smaller value of
the Hubble parameter. A more thorough discussion is given
in ref. \cite{S95}.

\subsection{Discussion with Grishchuk}

In this conference, Grishchuk \cite{G94} has stated that
he believes that the tensor fluctuations arising
from virtually all inflation models will dominate the
contribution to the microwave
background anisotropy observed by COBE. Unfortunately,
I disagree with this claim. I feel that he
has not adequately treated heating of the Universe after
inflation. In my Ph.D. thesis
(see ref. \cite{SBB89}), I computed the evolution of the
metric fluctuations during the transition from the
inflationary epoch to the radiation-dominated era
for a phenomenological model. Although the maximum
temperature of the radiation varies sensitively
on the damping coefficient $\Gamma$, there was basically no effect
on the scalar perturbations: $\zeta$ was constant in time.
More precisely, on a comoving
slice where the matter is at rest, the fluctuations in the metric,
(both scalar and tensor) were constant. These metric fluctuations
were essentially dormant whenever the physical wavelength of a given mode
exceeded the Hubble radius. The anisotropies
we observe now in large angle microwave experiments have
basically survived intact since the modes left the Hubble radius
during inflation.

\section{APPLYING THE ZEL'DOVICH APPROXIMATION TO GENERAL RELATIVITY}

In a Newtonian framework, Zel'dovich \cite{Zel} showed how
to extrapolate the results of cosmological perturbation theory
for a matter-dominated Universe
to the quasilinear regime. Recently, we
(Croudace, Parry, Salopek and Stewart  ---
hereafter CPSS \cite{CPSS94}) have been
able to derive the Zel'dovich approximation
and its higher order corrections using
general relativity. The simplicity of our results
should prove to be a useful tool in cosmology.
Here, I will review how to use Hamilton-Jacobi methods
in solving Einstein's equations.
We hope to apply our techniques to the interpretation
of redshift surveys. An alternative approach
to incorporating the Zel'dovich approximation within
general relativity has been suggested by
Matarrese {\it et al} \cite{M93}; see also ref.\cite{ELLIS94}.

\subsection{Gradient Expansion of 3-metric}

We assume from the beginning that cold-dark-matter is described
by collisionless, dust-like particles whose 4-velocity,
$U^\mu$,  is given
by the 4-gradient of the potential  $\chi$:
\begin{equation}
U^\mu = -  \chi^{,\mu} .
\end{equation}
Throughout, we will observe the line-element on a
time-hypersurface where $\chi$ is uniform:
\begin{equation}
ds^2 = -d \chi^2 + \gamma_{ij}(\chi, q) dq^i dq^j   \, .
\end{equation}
This choice is referred to as comoving, synchronous
gauge, in which case the 3-metric $\gamma_{ij}$ has physical
significance:
it gives the physical distance,
$ds = (\gamma_{ij} dq^i dq^j)^{1/2}$,
between two
comoving observers separated by an infinitesimal
comoving distance $dq^i$.  The evolution equation
for the 3-metric then becomes
\begin{equation}
{ \partial \gamma_{ij} \over \partial \chi }
=  2  \gamma^{-1/2}\,
{ \delta { \cal S}\over \delta \gamma_{kl}(x)}\,
  \biggl ( 2  \gamma_{ik} \gamma_{jl} - \gamma_{ij} \gamma_{kl}
  \biggr ) \, ,  \label{EVOL}
\end{equation}
where the generating functional ${\cal S}$ is given
by an expression analogous to eq.(\ref{theexpansion}).

During the matter-dominated era, the peculiar velocity
of the matter is typically small compared to the speed of light.
Hence we are justified in assuming that all spatial gradients
are small compared to unity,
\begin{equation}
R \ll 1 \ \,,
\end{equation}
where $R$ is the Ricci curvature of the 3-metric.
In fact, if we retain only the long-wavelength piece,
${\cal S}^{(0)}$ (analogous to eq.(\ref{LW})),
in the evolution equation (\ref{EVOL}),
we determine that the 3-metric evolves according to
\begin{equation}
\gamma^{(1)}_{ij}(\chi, q) = a^2(\chi) \, k_{ij}(q) \, , \ \
{\rm where} \  \ a(\chi) \equiv  \chi^{2/3}
\end{equation}
which is accurate to first order in spatial gradients.
The `seed metric' $k_{ij}(q)$ is an arbitrary function
that is independent of time;
it describes the initial fluctuations whose wavelength
is larger than the Hubble radius. This result is very
old --- it was known to Lifshitz and Khalatnikov \cite{L64}
(see also Tomita \cite{T75}).

One may solve for higher order terms by adopting
an iterative method. For example, accurate to
third order in spatial gradients, we use eq.(\ref{S2A})
to find
\begin{equation}
\gamma^{(3)}_{ij}(\chi, q) =
a^2(\chi) \left [  k_{ij}(q) +
{ 9 \over 20 } a(\chi) \,
\left (  \hat R(q) \,  k_{ij}(q) - 4 \hat R_{ij}(q)   \right )
\right ]  \ \, \quad {\rm (3rd \ order)} \label{third}
\end{equation}
where $\widehat R_{ij}$ is the Ricci tensor of the
seed metric $k_{ij}$.

Unfortunately, after a sufficient amount of time,
the third order expression, eq.(\ref{third}),
leads to nonsensical results: the determinant of the
3-metric can actually become
negative.  A similar problem occurs when one naively
approximates a non-negative function ${\rm cos}^2(x) \sim 1- x^2$
with the first two terms of a
Taylor expansion --- the approximate function falls below
zero when $x > 1$. One can easily remedy this problem by expanding
${\rm cos}(x) \sim 1- x^2/2$ , and then approximating ${\rm cos}^2(x)$
by the square of this result --- this technique guarantees
a positive result. In an analogous way,
we can improve the expansions eqs.(5,6) by expressing the results
as a `square.' The improved 3-third order result
\begin{eqnarray}
\tilde \gamma^{(3)}_{ij}(\chi,q) =
a^2(\chi)
&&\left \{ k_{il} + { 9 \over 40 } a(\chi)
\left [ \hat R k_{il} - 4 \hat R_{il} \right ] \right \} \,
k^{lm} \, \\
&&\left \{ k_{jm} + { 9 \over 40 } a(\chi)
\left [ \hat R k_{jm} - 4 \hat R_{jm} \right ] \right \}
\label{improved}
\end{eqnarray}
yields the relativistic generalization of the
Zel'dovich approximation.

In Fig.(6), we compare our approximations with an
exact Szekeres solution with azimuthal symmetry
\cite{SZ75}. It is remarkable that the
improved 3rd order expansion yields the exact result.
Higher order corrections to the Zel'dovich
approximation are discussed in
refs.\cite{CPSS94}, \cite{SSCP94}, \cite{SSC94}.

\section{ONE-LOOP APPROXIMATION APPLIED TO INFLATION}

This field is still in its infancy, but several notable
steps have already been taken. I will focus my
attention on curvature-coupled models, but I will also mention
other avenues that are being pursued.

One of the main problems with inflation is that there
are too many models. Some of these are very heavily
constrained by observations, but can theoretical arguments
be used to restrict the models further?

In the standard model of particle physics with energies
below 1 TeV, renormalizability arguments proved to very
effective in limiting the class of permissible Lagrangians.
For example, in the Higgs sector, one considers only those scalar field
potential which were polynomials of degree 4 or less, otherwise
the theory is not renormalizable.
In an analogous way, it would be very useful to have a
renormalizable theory of inflation. However, this is quite a
daunting task because this would require a
quantum theory of the gravitational field! Many
great minds have been wrecked on these treacherous
shores, so we will have to proceed cautiously.

I would say that our present
understanding of inflationary theory is comparable to that of
weak interactions when Fermi introduced
his 4-fermion theory in the 1930's.  Fermi's theory
was an outstanding example of phenomenological work,
but only thirty years later in the late 1960's was a renormalizable
theory of the weak interactions formulated.
For the same reason, I suspect that it may be several decades
before we obtain a satisfactory quantum formulation of the
gravitational field, whether it be a renormalizable theory
or even a finite theory as suggested by superstrings
(see, e.g., ref.\cite{Gasp91}, \cite{CAMPBELL91}).

HJ theory, which is accurate to lowest order in
Planck's constant ($\hbar^0$) has
proven to be a very effective, but
can we expand to higher order in $\hbar$?
For instance, expanding the wavefunctional to next order
in $\hbar$  corresponds to the one-loop approximation.
Some partial progress has been made in this area provided
one considers only matter fields running in the loops.

  \subsection{Economical Formulation of Cosmological Inflation}

It is possible to construct an economical model of
cosmological inflation where the GUT Higgs is identified
with the inflaton which is also the Brans-Dicke scalar.
One begins with the action for the 4-metric $g_{\mu \nu}$
interacting with a scalar field $\phi$:
\begin{equation}
{\cal I} = \int \sqrt{ -g} \left \{
\left [ { m^2 \over 16 \pi }
- {1 \over 2 } \xi \phi^2 \right ] \; {}^{(4)}R
- { 1 \over 2 } g^{\mu \nu } \phi_{,\mu} \phi_{,\nu}
- { \lambda \over 4 } ( \phi^2 - \sigma^2 )^2 \right \} \, .
\label{ACTION}
\end{equation}
The $ - \xi \, {}^{(4)}R \phi^2 / 2$ term must appear in order to
cancel infinities when one considers a
$\lambda \phi^4/4$ self-interaction in curved space-time.
This theory is relatively easy to understand.
The coefficient of the Ricci 4-curvature ${}^{(4)}R$ can
be interpreted in terms of the
{\it effective} value of the Planck mass
\begin{equation}
G^{-1}_{eff}(\phi) \equiv m_{ {\cal P} \, eff}^2(\phi)
\equiv m^2 - 8 \pi \xi \phi^2 \,.
\label{GEFF}
\end{equation}
Here $m^2$ is the bare contribution to $m_{ {\cal P} \, eff}^2$
whereas
$- 8 \pi \xi \phi^2$ is the Higgs contribution. In order that
the Universe inflate sufficiently, one requires that \cite{SBB89}
\begin{equation}
\xi < 0.002 \,  \quad {\rm if} \quad  m \sim m_{\cal P}\, .
\label{LIMIT}
\end{equation}
Although this is a rather crude constraint, it is nonetheless
very vexing for chaotic inflation.
If this constraint is violated, Newton's
constant $G_{eff}$ can become negative. Classically, it would have to
remain negative because it must cross a singularity in order to reach
positive values.

    \subsubsection{Variable Planck Mass Model and Induced Gravity}

The simplest way to satisfy eq.(\ref{LIMIT}) is to take
$\xi$ to be negative. Then there are essentially two cases to consider. In
{\it Induced Gravity}, one legislates that the bare value $m$ vanish, and
that the present value of the Planck scale is determined when the
scalar field rolls to its
minimum $\sigma$ in the potential. This is a particularly attractive model
requiring the fewest number of parameters, although it may not
produce a radiation-dominated Universe after inflation
(which would be a disaster). In the second case which is adopted here
one assumes that $m \sim m_{\cal P}$, and that the present value of the
Planck scale obtains
only a small contribution from $\phi$ at the minimum of its potential,
although it was much much larger during the inflationary era when
$\phi >> \sigma$. This situation is called the
{\it Variable Planck Mass Model}.

Since the gravitational sector of eq.({\ref{ACTION}) is a
bit messy, it proves
very convenient to employ a conformal transformation of the metric
\begin{equation}
g_{\mu \nu } = \Omega^2(\phi) \tilde g_{\mu \nu } \, , \quad
\Omega(\phi)  = m_{\cal P} / m_{ {\cal P} \, eff}(\phi) \,.
\end{equation}
This transformation suggested by Brans and Dicke \cite{BD61} is easy to
interpret--- it is just a change of units from a variable ruler
$m_{ {\cal P} \, eff} (\phi)$ to a uniform ruler $m_{\cal P}$ (which is
the current value of the Planck scale). Much of the physics
that is discussed below is similar to changing units in the same
way that one converts from feet to metres. The new Lagrangian density
\begin{equation}
\tilde {\cal L} = { m_{\cal P}^2 \over 16 \pi } {}^{(4)}\tilde R
- { 1 \over 2 } \tilde g^{\mu \nu } \chi_{,\mu} \chi_{,\nu}
  - V_{eff}(\phi(\chi))
\end{equation}
describes standard Einstein gravity for the new metric
$\tilde g_{\mu \nu }$. The effective potential given by
\begin{equation}
V_{eff}(\phi) = \Omega^4(\phi) \, V(\phi)
\end{equation}
is very flat as $ |\phi| \rightarrow \infty$ which is very
desirable for
inflation. $\chi$ is a function of $\phi$, and it is given
in ref.\cite{SBB89}. For induced gravity, its expression
is quite simple
\begin{equation}
\phi= \sigma \, e^ { \alpha \chi }\, , \quad
V_{eff}(\chi) =\; { \lambda \over 4} \;
({ m_{\cal P}^2 \over 8 \pi \vert  \xi\vert  }
)^2 \; ( 1 - e^{-2\alpha \chi} ) ^2 \, , \quad
\alpha =
\Big( {8 \pi \vert \xi\vert  \over 1 + 6 \vert \xi\vert  } \Big)^{1 / 2}
\, m_{\cal P}^{-1} \, .
\end{equation}
The effective potential for both models is plotted in Fig.(7).

The fluctuation spectrum for the Variable Planck Mass model
is essentially the same as for induced gravity:
\begin{equation}
{ \cal P}_{\zeta}(k) =  { 1 \over  8 \pi^2  }
 { \lambda \over \xi^2 } N^2_I(k) \, , \quad {\rm where}
\quad N_I(k) =
60 - \ln \left [ k/ ( 10^{-4} {\rm Mpc}^{-1}) \right ] \, .
\end{equation}
Normalizing according to the COBE observations, one obtains
\begin{equation}
{ \lambda \over \xi^2} = 3.9 \times 10^{-10} \pm 18\%\, \quad
b_\rho= 0.91 \pm 9\% \, .
\end{equation}
Gravity waves contribute about 0.3\% to $\sigma^2_{sky}(10^0)$.
If $\phi$ is the GUT Higgs, then $ \lambda \sim 0.05$ is
consistent with radiative corrections and one can fit COBE if
$\xi \sim - 10^4$. The spectral index for scalar fluctuations
is very close to unity, $n = 0.967$. In order to reconcile
this result with galaxy data, one would have to alter
the standard cold-dark-matter transfer function.
For example, a mixed-dark-matter model \cite{MDM}
could be invoked to explain Fig.(2).

The main advantage of having large negative $\xi$ is
that one-loop radiative corrections do
not destroy the required flatness in the effective
potential \cite{SBB89}.
For example, one can couple a gauge boson $A_\mu$ through the
minimal prescription replacing
\begin{equation}
\partial_\mu \rightarrow \partial_\mu  - ie A_\mu \, ,
\end{equation}
and by adding kinetic terms $ F_{\mu \nu } F^{\mu \nu }$
to the original action (\ref{ACTION}). After the conformal
transformation is
performed, the gauge field $A_\mu$ does not transform---
it is conformally invariant. Its mass is given by the
Higgs mechanism
\begin{equation}
m_A(\phi) = e \phi \, \Omega(\phi)
\end{equation}
which is scaled by  the additional factor $\Omega(\phi)$.
If $ m \ne 0$, then for small values of
$\phi$, one recovers the usual coupling expression, but
for large values of $\phi$, the gauge boson mass is independent
of $ \phi$: the gauge boson decouples from the Higgs field.
As a result the 1-loop radiative potential \cite{SBB89}
\begin{equation}
V_{rad}(\phi) = { 3 \over 64 \pi^2} m_A^4(\phi) \, \ln
\left ( m_A^2(\phi) \over \mu^2 \right )
\end{equation}
remains very flat as $ \phi \rightarrow \infty$ ($\mu$ is the
renormalization scale.)
The above argument works for fermions as well. By altering the
gravitational sector, one has a robust method of producing a flat
potential that proves favorable for the inflationary scenario.

For induced gravity with
$m=0$, the mass of the gauge boson
$m_A= e m_{\cal P}/ \sqrt { 8 \pi |\xi|}$ is independent
of the scalar field, and hence the scalar field does not couple directly to
radiation. Hence, there may be a problem with producing a
radiation-dominated era at the end of inflation.

However, for the Variable Planck Mass model, heating is
efficient, and the maximum temperature
\begin{equation}
T_{max}
\sim \epsilon \, \left ( { 15 \over 128 \pi^4} g_{eff}^{-1}
 { \lambda \over \xi^2} \right  )^{1\over 4}
m_{\cal P}= 1.2 \times 10^{-4} m_{\cal P} \, ,
\end{equation}
reached after inflation is rather high
and in order not produce magnetic monopoles one requires that
the GUT scale satisfy
\begin{equation}
\sigma >   1.5 \times 10^{-4} m_{\cal P} \, .
\end{equation}
This result is computed for the SU(5) model, and it is
slightly model dependent (the effective number of
degrees of freedom was taken to be $g_{eff}=160.75$ and
the efficiency factor for heating was $\epsilon \sim 0.5$).

  \subsection{Quantum Scale of Inflation}

In the context of quantum cosmology,
Barvinsky and Kamenshchik \cite{BK94}
considered the Variable Planck Mass model
using a 1-loop approximation. They point
out that the initial value of the scalar field
is a free parameter which can be constrained
using arguments of quantum cosmology.

At tree-level, they find that the wavefunction
for the scalar field  in a minisuperspace model
is not normalizable. However,
in a 1-loop approximation, they find that it
is strongly peaked with values that are
consistent the measurement of the microwave
background anisotropy. This is an interesting
development for quantum cosmology, and it should
be pursued further. Moreover, the
wavefunction of the Universe can actually
be measured \cite{S92}. For the latest experimental
results concerning the statistics of the
large angle microwave background measurements,
consult Kogut \cite{K94} who was able to rule
out some toy non-Gaussian models.

  \subsection{Other One-Loop Effects}

Unfortunately, space limitations do not permit
a detailed discussion of other
one-loop effects in inflation. For completeness,
I will mention the main areas:

\noindent
(1) stochastic inflation
\cite {Starobinsky}-\cite{Linde2} ----
in this proceedings,
Linde \cite{Linde94} has given an enthusiastic discussion;

\noindent
(2) heating of the Universe after inflation is
an interesting topic \cite{DeVega94};

\noindent
(3) $R^2-$inflation \cite{STAROBINSKY83},
\cite{S92}, \cite{SBB89} ---- here, one-loop effects
drive inflation;

\noindent
(4) computation of the curvature coupling constant
$\xi$ ---- consult Hill and Salopek \cite{HS90} as well as
Buchbinder {\it et al} \cite{BUCH89} and
Reuter \cite{REUTER94};

\noindent
(5) generation of primordial magnetic fields from inflation
through the breaking of conformal invariance
(Dolgov \cite{DOLGOV93});

\noindent
(6) string field theory and inflation ----
whether string theory  is compatible with inflation
is still an open question. In this volume,
related issues are discussed by Norma
Sanchez \cite{SANCHEZ94}. See also ref.\cite{CAMPBELL91}.

I suspect that the next major stage in the
development of inflation theory will be
a careful treatment of one-loop effects.
Although, this subject is technically quite challenging,
many problems are indeed tractable.
I feel that the fascinating theory
of one-loop effects on inflation will be of interest
for many years to come.

\section{CONCLUSIONS}

The inflationary scenario provides a relatively
good understanding of many cosmological observations,
including galaxy clustering and microwave anisotropies.
As the data continues to improve, one hopes to
determine the correct inflationary model.

In order to facilitate the comparison of models
with observations, some very powerful theoretical
machinery has been developed. In fact, Hamilton-Jacobi
theory provides an elegant means to solve
many problems appearing in cosmology.

Here I have reviewed how to compute galaxy correlations
and microwave anisotropies arising from
power-law inflation where the scale factor evolves
according to $a(t) \propto t^p$. This model helps to alleviate
the problems of large scale structure. For example,
if $p=21$, the spectral index for scalar perturbations is
found to be $n=0.9$. This models yields a $35 \%$ gravity wave
contribution to $\sigma^2_{sky}(10^0)$ (determined by COBE).
Although the fit to large scale clustering is not
perfect, I feel that with improvements in the
quantity and quality of data, one should obtain a better fit.

Another nice application of Hamilton-Jacobi
theory is incorporating the Zel'dovich approximation
within general relativity. The Zel'dovich approximation
describes the formation of sheet-like structures which
appear to be quite common in our Universe.

At the moment, the future for the inflationary scenario
appears quite bright. Already researchers are considering
going beyond the semi-classical approximation provided by
Hamilton-Jacobi theory: they are now investigating
one-loop corrections to inflation.

\section{ACKNOWLEDGMENTS}

I would like to thank Norma Sanchez for organizing
an enjoyable conference. Some of the work that was
reviewed here was done in collaboration with John Stewart
at Cambridge University. I acknowledge partial support
from the Natural Sciences and Engineering Research Council
of Canada, and the Canadian Institute for Theoretical
Astrophysics in Edmonton.

\vfill\eject
\section{Figure Captions}

\noindent
{\bf Fig.(1)}:COBE has measured the angular correlation function
$C(\alpha)$ for the temperature anisotropy with the monopole and dipole
removed. Below about 60 degrees, the measurements are clearly
much larger than the error bars. In fact, COBE claims a
10 standard deviation detection. (Taken from Bennett et al [5].)

\noindent
{\bf Fig.(2)}: The observed galaxy power spectrum
is shown. For comparison, the 1.8 power-law
that is usually associated with optical galaxies
is also shown.

\noindent
{\bf Fig.(3)}: A plot of recent microwave background measurements
is shown. Since one is observing the celestial sphere,
it is useful to decompose into spherical harmonics,
$Y_{\ell m}$, where $\ell$ is harmonic the index, and
$T_{\ell}^2$ is the average amplitude for a particular
$\ell$. The solid curve depicts the standard cold-dark-matter
model prediction. For $\ell > 20$, there is much scatter
in the observational data, and more data are required.
(Taken from Wright et al [21].)

\noindent
{\bf Fig.(4)}: Primordial scalar perturbations of the metric
are described by the function $\zeta$. The power spectra for
zeta are shown
for various choices of the the spectral index $n= 1- 2/(p-1)$
arising from power-law inflation. They have been normalized using
COBE's 2-yr data set.

\noindent
{\bf Fig.(5)}: For the present epoch, power spectra
for the linear density perturbation $\delta \rho/ \rho$
are shown for various cold-dark-matter (CDM) models.
The points with error-bars are the observed data
with $b_{IRAS}=1$. Standard CDM utilizes a Zel'dovich
spectrum with a spectral index $n=1.0$. It
is unsatisfactory at short scales.
The dark line depicts the best fit of the power-law inflation
models with $n=0.9$. It yields a 35\% gravitational wave
contribution to the large-angle microwave background anisotropy
$\sigma^2_{sky}(10^0)$.

\noindent
{\bf Fig.(6)}: The various orders of the gradient expansion
are compared with the exact Szekeres solution
(bold curve) for the evolution of the
3-metric component $\gamma_{33}$ in terms of
the scale factor $a(\chi) \equiv \chi^{2/3}$.
Pancake formation occurs when $\gamma_{33}=0$.
The thin curve is the first order
term (long-wavelength), whereas the dotted graph
is the third order result.
The improved third order expansion
yields the exact result.

\noindent
{\bf Fig.(7)}: Inflation can be reconciled with
particle physics using the (b) Variable Planck Mass model
which is closely related to (a) Induced Gravity.
The effective potentials are plotted as a
function of $\chi\equiv \chi(\phi)$. For large positive values
of $\chi$, the effective potential is extremely flat as
required by inflation.
One-loop radiative corrections do not destroy the flatness of the
potential.

\end{document}